\documentclass[a4paper,12pt]{article}
\pdfoutput=1
\usepackage{graphicx, rotating, slashed, color, colortbl}
\usepackage{graphicx, rotating, color}

\usepackage[usenames,dvipsnames]{xcolor}
\usepackage{graphicx}
\usepackage{colortbl}
\usepackage[unicode=true,pdfusetitle,
 bookmarks=true,bookmarksnumbered=false,bookmarksopen=false,citecolor=Turquoise,linktoc=page,
 breaklinks=false,pdfborder={0 0 1},backref=false,colorlinks=true,pdfpagemode=FullScreen]
 {hyperref}
\usepackage{amssymb}

\newcommand{\mev}{\textrm{ MeV}}
\newcommand{\gev}{\textrm{ GeV}}

\newcommand{\vev}[1]{\langle #1 \rangle}

\newcommand{\arxiv}[1]{ \href{http://arxiv.org/abs/#1}{ #1}}

\usepackage{color}
\definecolor{rosso}{cmyk}{0,1,1,0.4}
\definecolor{rossos}{cmyk}{0,1,1,0.55}
\definecolor{rossoc}{cmyk}{0,1,1,0.2}
\definecolor{blu}{cmyk}{1,1,0,0.3}
\definecolor{blus}{cmyk}{1,1,0,0.6}
\definecolor{bluc}{cmyk}{1,1,0,0.1}
\definecolor{verde}{cmyk}{0.92,0,0.59,0.25}
\definecolor{verdec}{cmyk}{0.92,0,0.59,0.15}
\definecolor{verdes}{cmyk}{0.92,0,0.59,0.4}
\definecolor{grigio}{cmyk}{0,0,0,0.1}
\definecolor{rosa}{cmyk}{0,0.1,0.1,0.02}
\definecolor{rosino}{cmyk}{0,0.05,0.05,0.02}
\definecolor{rosas}{cmyk}{0,0.3,0.25,0.05}
\definecolor{celeste}{cmyk}{0.1,0,0,0.02}
\definecolor{giallino}{cmyk}{0,0,0.1,0.02}
\definecolor{rosso}{cmyk}{0,1,1,0.4}
\definecolor{rossos}{cmyk}{0,1,1,0.55}
\definecolor{rossoc}{cmyk}{0,1,1,0.2}
\definecolor{blu}{cmyk}{1,1,0,0.3}
\definecolor{blus}{cmyk}{1,1,0,0.5}
\definecolor{bluc}{cmyk}{1,1,0,0.1}
\definecolor{blucc}{cmyk}{0.7,0.5,0,0}
\definecolor{viola}{cmyk}{0,1,0,0.6}
\definecolor{viola2}{cmyk}{0,1,0.2,0.6}
\definecolor{verde}{cmyk}{0.92,0,0.59,0.25}
\definecolor{verdec}{cmyk}{0.92,0,0.59,0.15}
\definecolor{verdes}{cmyk}{0.92,0,0.59,0.4}
\definecolor{verdino}{cmyk}{0.12,0,0.09,0.02}
\definecolor{giallo}{cmyk}{0,0,1,0}
\definecolor{gialloverde}{cmyk}{0.44,0,0.74,0}

\usepackage[small]{caption}
\begin{document}
\begin{flushright}
UMD-PP-013-001\\ 
January 2013
\end{flushright}

\bigskip
\begin{center}
{\huge New Patterns of Natural R-Parity Violation with Supersymmetric Gauged Flavor\\} 
\vspace*{1cm}
{\bf \Large R. Franceschini and R. N. Mohapatra\\}
\vspace{0.3cm}
{ \large \it Maryland Center for Fundamental Physics and Department of Physics\\ University of Maryland, College Park, MD 20742, USA}

\bigskip\bigskip

\centerline{\large\bf Abstract}
\begin{quote}\large
\noindent We point out that supersymmetric gauged flavor models provide a realization of R-parity violation (RPV) that is natural in the sense that it does not lead to catastrophic proton decay for natural values of parameters in the theory. Within specific realizations of the idea, the relative strengths of the $\Delta B=1$ $u^cd^cd^c$ type RPV operators can be predicted.  In particular, we present examples of gauged flavor models where RPV couplings depend on quark masses as $(m_{u_i}m_{d_j}m_{d_k}/m^3_t)^n$ where $n=1$ or $n= 1/2$. Some phenomenological implications of these models are discussed.
\end{quote}

\end{center}

\newpage

\tableofcontents

\bigskip

\section{Introduction}
It is well known that the Minimal Supersymmetric Standard Model (MSSM) allows for the so-called R-parity violating (RPV) terms in its superpotential. Using the standard superfield notation, we have~\cite{hall}
\begin{eqnarray}
W_{RPV}~=~\lambda LLe^c +\lambda^\prime QLd^c +\lambda^{\prime\prime} u^cd^cd^c + \mu_L LH_u\,.
\end{eqnarray}
These terms do not conserve baryon and lepton number and the presence of all these terms can be quite undesirable: for instance, (i) when any of the terms by itself is present, it allows the lightest stable supersymmetric partner (LSP) to decay with implications for whether dark matter is a SUSY partner or not ; (ii) secondly, and more importantly, 
when the second and the third terms are present, they lead to instantaneous proton decay  unless the product of $\lambda^\prime \lambda^{\prime\prime} \leq 10^{-24}$ when first generation fermions are involved. Clearly this is an unnatural fine tuning of parameters.  It is also worth pointing out that non-negligible RPV couplings can significantly affect the current LHC bounds on super-partner masses and has been of interest to phenomenologists to see if SUSY breaking masses can be in the range which avoids excessive fine tuning to understand the weak scale. It is therefore of interest to seek extensions of MSSM where constraints of high scale physics can help to restrict the form of or completely eliminate the RPV couplings and in particular avoid the constraints of proton decay in a natural manner while keeping the strength of RPV couplings to be same  order of the quark Yukawa couplings, so that they can have a role in colliders studies.

Examples of theories exist in literature where the form of the RPV interactions is naturally restricted. For instance, a generic framework, where RPV do not lead to proton decay is the one where R-parity violation is spontaneous~\cite{aulakh} by the vacuum expectation value of fields such $\tilde{\nu}$ in MSSM or extended models with right handed sneutrinos~\cite{kuchi,spinner}. 
For realistic models of this type, see \cite{kuchi,spinner} where the pure hadronic RPV coupling $\lambda^{\prime\prime}$ vanishes, thereby eliminating the threat of proton decay.
In this paper, we seek alternative frameworks where the $\Delta B=1$ $\lambda^{\prime\prime}$ coupling may be present but not the $\lambda^\prime$ term, so that again there is no threat of proton decay. We hasten to add that in supersymmetric models which have a super-light gravitino, we do expect $p\to K^{+} \tilde{G}$ with life times easily compatible with current lower limits (and therefore no threat of proton decay as emphasized).  A class of models which seem to have this property are the so-called gauged flavor models, which is the focus of this paper.

Gauged flavor models start with the idea of using the flavor symmetries (or a subgroup) of the standard model in the limit of zero fermion masses as local symmetries that supplement the standard model electroweak symmetry or its extensions . A conceptually attractive aspect of these models is that flavor patterns of quarks and leptons are supposed to arise as a dynamical consequence of breaking of horizontal (or flavor) symmetries via minimization of the flavor Higgs potential. To test the viability of this idea, one must either directly look for the flavor gauge bosons in colliders or look for their indirect effects in rare low energy processes. In a large class of models of this type, however, the flavor scale is pushed so high due to existing constraints from flavor changing neutral current effects (FCNC) that it makes it hard to test the idea experimentally.  Recently, however, it has been pointed out~\cite{grv} that if there are new vector-like fermions
at the TeV scale, then, the full flavor group of SM can be an anomaly free gauge symmetry with at least some flavor gauge bosons and new fermions within the reach of colliders. The reason for this is that the fermion masses arise via a quark analog~\cite{qseesaw}  of the seesaw mechanism
which leads to vector-like fermions having an  mass hierarchy inverted with respect to that of the SM fermions. Therefore the top partner is the lightest of the vector like fermion with mass close to a TeV. 
The gauged flavor model was extended to the case of left-right symmetric (LRS) electroweak group~\cite{lrs}  based on the group $SU(2)_L\times SU(2)_R\times U(1)_{B-L}$  \cite{gms} in order to naturally  accommodate neutrino masses and mixings. The LRS version has also the advantage that all masses in the theory owe their origin to gauge symmetry breaking and are therefore connected to other physics. 
Here we focus  on possible extensions of the left-right gauged flavor scenario to
include supersymmetry and study its consequences for baryon and lepton number violation.

 The first point we emphasize is that gauged flavor models lead to ``safe'' R-parity violation because the $QLd^c$ coupling is naturally forbidden due to separate lepton and quark flavor group~\cite{GFRPV}. Secondly, we point out that depending on the model, the RPV coupling $\lambda^{\prime\prime}$ can be suppressed by specific powers of quark mass ratios, which then not only provides a natural suppression of the RPV interactions but also a way to distinguish between various models of natural RPV. For the sake of illustration, we explore this question within the framework of left-right symmetric gauge models although we comment on the MSSM case in the beginning.

\bigskip

This paper is organized as follows: in Section~\ref{GFMSSM} we recall the nature of RPV in the gauged flavor version of MSSM; in Section~\ref{SUSYLRS} we discuss two different realization of gauged flavor in SUSY models with a gauge left-right symmetry and show how the strength of RPV couplings of $\Delta B=1$ pure quark and three lepton type depends on quark and lepton masses. In Section~\ref{pheno} we discuss the phenomenology of the presented RPV SUSY models with left-right gauge symmetry. In particular we focus on the observables that are sensitive to the structure of the RPV couplings that reflect the distinctive properties of the models.


\section{MSSM with gauged flavor and R-parity breaking \label{GFMSSM}}
The generic class of models we consider in this paper are supersymmetric electroweak models extended by the addition of vector-like fermions denoted by $U, U^c; D, D^c; E, E^c; N, N^c$.  The presence of the latter fermions allow full gauging of the flavor group that arises in the limit of zero fermion masses. 

In this section, we consider the first example of such models - the gauged flavor version of the MSSM.
In addition to the chiral superfields $Q, L, u^c, d^c, e^c, H_{u,d}$ which define the MSSM, we add three right handed neutrinos $\nu^c$ to accommodate neutrino masses. This extends the minimal gauged flavor model of Ref.~\cite{grv}.
The flavor group in this case is $$G_f\equiv SU(3)_Q\times SU(3)_{u^c}\times SU(3)_{d^c}\times SU(3)_{e^c}\times SU(3)_{\nu^c}$$ {and it is completely broken by the VEV of a set of Higgs fields that we call  flavons $Y_{u},\bar{Y}_{u},Y_{d},\bar{Y}_{d}$ and $Y_{\nu},\bar{Y}_{\nu},Y_{\ell},\bar{Y}_{\ell}$}. {The electroweak part of the SM gauge group is broken by a combination of the Higgs fields $H_{1}$ and $H_{2}$ that takes the VEV $v$. } 
{To cancel the anomalies of the flavor group $G_{f}$,} we add the vector-like {\it iso-singlet} fields $U, U^c, D, D^c, E, E^c; N,N^c$. The assignments of the  fields to representations of $G_f$ are given in Table~\ref{tab:charges}.
\begin{table}
\small
\begin{center} 
$$\begin{array}{|| l |ccc|cccccc||}\hline\hline
\rowcolor[gray]{0.9} &  SU(3)_c &  SU(2)_L  &   U(1)_Y   &  SU(3)_{Q}  &  SU(3)_{u^c}  &  SU(3)_{d^c}   &  SU(3)_{L}  &  SU(3)_{e^c}  &  SU(3)_{\nu^c} \cr
\hline \hline
\rowcolor{verdino}   Q & 3 & 2 &  \frac{1}{3}   & 3  & &  & & & \cr
\rowcolor{verdino}   u^c  &  3^{*}&  &  -\frac{4}{3}       &   & 3^{*} &     & & &\\ [3pt]
\rowcolor{verdino}     d^c & 3^{*} &  &  \frac{2}{3}       &   &           & 3^{*}    & & & \\ [3pt]
 \rowcolor{giallino}   U  & 3 &  &  \frac{4}{3}                  &   & 3 &  & & &\\ [3pt]
 \rowcolor{giallino}    U^c &   3^* &  &  -\frac{4}{3}      & 3^{*}  &    &    & & &\\ [3pt]
 \rowcolor{giallino}    D  &  3 & &  -\frac{2}{3}               &   &     & 3   && & \\ [3pt]
 \rowcolor{giallino}    D^c  & 3^{*} &  &  \frac{2}{3}     &  3^{*}& &  & & &\\ [3pt]
\hline
\rowcolor{verdino}    L  && 2  &  -1   & && & 3 & &  \\ [3pt]
\rowcolor{verdino}    e^c  && & 2   & & &   & & 3^*   &\\ [3pt]
\rowcolor{verdino}    \nu^c  && & 0  & & &   & && 3^*   \\ [3pt]
 \rowcolor{giallino}    E  &    &&  -2   & & &   & &3& \\ [3pt]
 \rowcolor{giallino}    E^c  &&    &  2   & && &  3^*  & &  \\ [3pt]
 \rowcolor{giallino}    N  &  &  & 0   & & &   & &&3 \\ [3pt]
 \rowcolor{giallino}    N^c  & &   & 0   & & &  &  3^*  & &  \\ [3pt]
\hline
  \rowcolor{verdino} H_u  & & 2 & 1  & &  & &&&\\ [3pt]
  \rowcolor{verdino} H_d  & & 2 & -1   &  & &&&&\\ [3pt]
   Y_u  &  &  & &  3  & 3^{*} & &&& \\ [3pt]
   \bar{Y}_u  &  &  &   &{3^{*}}  &   3  & &&& \\ [3pt]
   Y_d  &   &  &   &3  & &  {3^{*}}    &&&\\ [3pt]
   \bar{Y}_d  &  & &  &    {3^{*}}  & & 3  &  & &\\ [3pt]
   Y_\ell  & & & & & & & {3}  &   3^{*}  &\\[3pt]
   \bar{Y}_\ell  & & & & & & & {3^{*}}  &  3 &\\[3pt]
 Y_\nu  & & & & & &&  3  &  &3^{*}\\[3pt]
 \bar{Y}_\nu  & & & & & && {3^{*}}  &  &  3 \\[3pt]
\hline\hline
\end{array}$$
\end{center}
\caption{Particle content of the MSSM extension to accommodate the gauging of the flavor group SU(3)$^{6}$. The MSSM fields are the lines with green background, the new matter fields are given in the yellow lines, and the flavor symmetry breaking sector is given in the lines with white background. }
\label{tab:charges}
\end{table}
\normalsize


This model has the following RPV couplings
\begin{eqnarray}
W_{RPV}&=& \epsilon^{ijk}  \left[\lambda_q U^{c}_{i}D^{c}_{j}D^{c}_{k} + \lambda_1 {N^c_iN^c_jN^c_k}+\lambda_2 {N_iN_jN_k}+\lambda_3 {\nu^c_i\nu^c_j\nu^c_k}\right]\,,
\label{eq:HDObarionicRPV}
\end{eqnarray}
where $i,j,k$ are flavor indexes. We remark that these RPV couplings involve only the states that we added usual MSSM particle content.

In a fashion similar to the model of Ref.~\cite{daniel} we can formulate a seesaw model where $W \supset \lambda_{u} Q  H U^{c}  + \lambda^{\prime}_{u} U U^{c} Y_{u}+ \mu_{u} U u^{c}$ and similar terms for the down-type quarks and the leptons.
Then the light up-type quark masses are given by the generic formula: $$m_{u,i}\sim \frac{\lambda_{u} \mu_{u} v^2}{ {\lambda_{u}^{\prime}} \langle\hat{Y}_u \rangle}\,,$$ where  one can take $\langle\hat{Y}_u \rangle=Y_u V^{\dagger}_{CKM}$ thanks to a suitable flavor gauge transformation. The latter formula highlights the inverse proportionality of the mass of the quark and that of its exotic partner, hence justifying calling this scenario a seesaw model. The striking consequence of this observation is that  the partner of the up quark is expected to be heavy while the partner of the top quark should be much lighter.  The mixing between the light and heavy eigenstates of the  mass matrix is $\theta_L \sim \frac{\lambda_{u} v}{\lambda_{u}^{\prime} \langle \hat{Y}_u \rangle}$ for left handed quarks and
$\theta_R \sim \frac{\mu_{u}}{\lambda_{u}^{\prime}\langle \hat{Y}_u\rangle}$ for right handed quarks {\cite{grv}. 

From eq.~(\ref{eq:HDObarionicRPV}) and the mixing angles one can express the strength of the generated RPV couplings in terms of the masses of the quarks. The dominant RPV couplings for the quarks is given by the generic formula:
\begin{eqnarray}
\lambda^{\prime\prime}_{123}\sim  \frac{m_sm_d}{m^2_t}\,.
\end{eqnarray}
This result is similar to the predictions obtained using the MFV ansatz~\cite{yuval}.

\section{SUSYLRS with gauged flavor \label{SUSYLRS}}
\subsection{Model I}
When the MSSM is extended to a SUSY left-right symmetric model, due to the presence of a local $B-L$ symmetry, the form of RPV interactions is severely restricted and how RPV finally manifests depends on the details of the matter content and symmetry breaking. This is because R-parity  is related to $B-L$ as $R=(-1)^{3(B-L)+2S}$. In fact it is known~\cite{martin} that if the Higgs sector of the model is such that the $B-L$ symmetry is broken by two units, then RPV interactions are absent to all orders. In this section, we discuss
the implications of SUSY left-right models with gauged flavor for the nature of RPV interactions.

The gauged flavor with left-right symmetric electroweak interactions  but without supersymmetry is discussed in Ref. \cite{gms}  {from which we borrow the notation.} 
Here we consider the supersymmetric version of this model. Some aspects of this model were noted in Ref.~\cite{GFRPV}.
The largest flavor group for this case is $SU(3)_{Q_L} \times SU(3)_{Q_R}\times SU(3)_{\ell_L}\times
SU(3)_{\ell_R}$.
{For the cancellation of anomalies we introduce vector-like superfields $U, U^c,D, D^c$ and $E, E^c,N, N^c$, analogously to the case of the MSSM.}
{The overall anomaly free gauge group is therefore}  $$G_{LR} \equiv
SU(3)_{c} \times SU(2)_L \times SU(2)_R \times U(1)_{B-L} \times
SU(3)_{Q_L} \times SU(3)_{Q_R}\times SU(3)_{\ell_L}\times
SU(3)_{\ell_R},$$ where $SU(3)_{Q_L} \times SU(3)_{Q_R}$
represents the flavor gauge symmetries respectively in
the left- and right-handed quark sector, and $SU(3)_{\ell_L}
\times SU(3)_{\ell_R}$ the corresponding ones for the lepton sector.
{The electroweak part of the SM gauge group is embedded into the group $SU(2)_{L}\times SU(2)_{R}\times U(1)_{B-L}$ that is broken to the SM group $SU(2)_{L}\times U(1)_{Y}$ by the VEV $v_{R}$ of a set of fields $\chi^{c},\bar{\chi}^{c}$.}
{The flavor symmetry is broken by a set of flavons $Y_{u},\bar{Y}_{u},Y_{d},\bar{Y}_{d}$ and $Y_{\nu},\bar{Y}_{\nu},Y_{\ell},\bar{Y}_{\ell}$ as in the example of the MSSM}.
 The fields of the model and their transformation properties under
 the group $G_{LR}$ are reported in Table \ref{tab:modelone}. 
 \begin{table}
\small
\begin{center} \begin{tabular}{|| l | c  c c c  | c c c c || }\hline\hline
&  $SU(3)_c$ & $SU(2)_L$ & $SU(2)_R$ & $U(1)_{B-L}$ & $SU(3)_{Q_L}$ &
  $SU(3)_{Q_R}$ &  $SU(3)_{\ell_L}$ & $SU(3)_{\ell_R}$ \\ [3pt]
\hline \hline
\rowcolor{verdino}  $Q$ & 3 & 2 & & $\frac{1}{3}$ & 3 &    & & \\ [3pt]
\rowcolor{verdino}  $Q^c$ & $3^*$ & & 2 & $-\frac{1}{3}$ &   & $3^*$  & & \\ [3pt]
   \rowcolor{giallino}  $U$ & 3 & & & $\frac{4}{3}$   &   & 3  & & \\ [3pt]
  \rowcolor{giallino}   $U^c$ & $3^{*}$ & & & $-\frac{4}{3}$   & $3^{*}$ &    & & \\ [3pt]
 \rowcolor{giallino}    $D$ & 3 & & & $-\frac{2}{3}$ &    & 3  & & \\ [3pt]
 \rowcolor{giallino}    $D^c$ & $3^*$ & & & $\frac{2}{3}$  & $3^*$ &   & & \\ [3pt]
\hline
\rowcolor{verdino}  $L$ & & 2 & & $-1$ & &  & 3 &   \\ [3pt]
\rowcolor{verdino}  $L^c$ && & 2 & $1$ & & &   & $3^*$ \\ [3pt]
   \rowcolor{giallino}  $E$ & &&   & $-2$ & & &   & $3$ \\ [3pt]
  \rowcolor{giallino}   $E^c$ && &   & $2$ &  & & $3^{*}$ &   \\ [3pt]
  \rowcolor{giallino}   $N$ && &   & 0  & & &    & 3 \\ [3pt]
 \rowcolor{giallino}    $N^c$ && &   & 0  & &  & $3^*$ &   \\ [3pt]
\hline
  $\chi,\bar{\chi}$ & & 2 & & $\pm 1$ &  &  & &\\ [3pt]
  $\chi^c,\bar{\chi}^c$ & & & 2 & $\pm 1$ &  &  & & \\ [3pt]
  $Y_u$ & &  &  &  & $3$ & $3^{*}$ &&\\ [3pt]
  $\bar{Y}_u$& &  &  &  & $ {3^{*}}$ &$3$ && \\ [3pt]
  $Y_d$ &  &&  &  & $3$ &  $3^{*}$  & &\\ [3pt]
   $\bar{Y}_d$ & &  &  &  & ${3^{*}}$ & $3$  &  &\\ [3pt]
  $Y_\ell$ & & & & & & &$3$ &  $3^{*}$\\[3pt]
  $\bar{Y}_\ell$ & & & & & & &${3^{*}}$ & $3$\\[3pt]
$Y_\nu$ & & & & & & &$3$ &  $3^{*}$\\[3pt]
$\bar{Y}_\nu$ & & & & & & &${3^{*}}$ & $3$\\[3pt]
\hline
\end{tabular} \end{center}
\caption{Model I matter content and transformation properties. The MSSM fields are the lines with green background, the new matter fields are given in the yellow lines, and the electroweak and flavor symmetry breaking sector is given in the lines with white background.}
\label{tab:modelone}
\end{table}
\normalsize

The interaction of the MSSM  and the exotic fields is given by the superpotential 
$$W_{I} = \lambda_{u}( Q\chi U^{c} + Q^{c}\bar{\chi}^{c} U) + \lambda_{d}( Q\bar{\chi} D^{c} + Q^{c}\chi^{c} D) + \lambda_{u}^{\prime}Y_{u}UU^{c} + \lambda_{d}^{\prime}Y_{d}DD^{c}\,, $$
where the equality of the couplings are dictated by unbroken L-R parity at the scale at which we write this superpotential. {This assumption about the L-R parity can be removed without affecting our conclusions, though getting more involved formulas. Thus we will discuss explicitly only the L-R parity symmetric case.}
{From these interactions the mass terms are generated once the flavon fields $Y$ and the Higgs fields $\chi$ take a VEV. To fix our notation we take $\vev{\chi^{c}}=\vev{\bar{\chi}^{c}}=(0,v_{R})$ and $\vev{\chi}=(0,v_{L})$ and $\vev{\bar{\chi}}=(0,\bar{v}_{L})$.
The flavons can be written in a basis where the $Y_{d}$ are diagonal by mean of a suitable flavor rotation, after which the $Y_{u}$ is fixed. Therefore we assume that the VEV of the flavons are such that  $\vev{Y_{d}}=\vev{\bar{Y}_{d}}=\vev{\hat{Y}_{d}}$ and $\vev{Y_{u}}=\vev{\bar{Y}_{u}}=V_{CKM}\vev{\hat{Y}_{u}}V_{CKM}^{\dagger}$ where hat denotes diagonal matrices.} 

As shown in  Ref.~\cite{gms} in this model the SM fermion masses are given by  a seesaw formula~\footnote{Here $v_L$ and $\bar{v}_L$ are the analogs of the VEV of $H_u$ and $H_d$ in MSSM and therefore their ratio gives the parameter tan$\beta$. The factor tan$\beta$ can be absorbed into the redefinition of $\lambda^\prime_d$ and we therefore do not display it below.}: 
\begin{eqnarray}
M_d\simeq {\lambda^2_d \bar{v}_L v_R \over \lambda_{d}^{\prime}} \langle\hat{Y}_d\rangle^{-1},\quad M_u\simeq  {\lambda^2_u v_L v_R \over \lambda^{\prime}_{u}} \cdot V^{\dagger}_{CKM} \langle\hat{Y}_u\rangle^{-1}V_{CKM}\,.
\end{eqnarray}
As in the model discussed above, this implies a mass hierarchy among the vector-like states inverted with respect to that of the SM states. Therefore  the top ``partner'', and possibly its SUSY partner, is predicted to be the lightest colored exotic fermion with masses in the TeV range depending on the scale of $SU(2)_R$ breaking.
{In general the light fermions are an admixture of the usual MSSM interaction eigenstates and the exotic fields with mixing angle that, in the approximation $v_{L} \ll v_{R} \ll \vev{Y}$, reads 
$$\theta_{L,R}^{\,(u,d)} \simeq \frac{\lambda_{u,d} \,v_{L,R}}{\lambda_{u,d}^{\prime}\vev{Y_{u,d}} }\,. $$}

{The FCNC effects generated by flavor gauge bosons of this model and the effect of the heavy fermions and their mixing in the properties of the EW gauge bosons} was analyzed  in \cite{gms} and it was shown that for the case where
the $W_R$ mass is TeV or lower, the lightest  flavor gauge boson can have masses in the TeV range as well, thus making the model testable at the LHC.

{The RPV couplings in this model are given by renormalizable interactions among the exotic states:}
\begin{eqnarray}\label{RP}
W_{RPV}~=~\lambda_q\epsilon_{ijk} \left[ U^c_iD^c_jD^c_k+U_iD_jD_k\right]+\lambda_\ell \epsilon_{ijk} \left[ N^iN^jN^k+N^{c,i}N^{c,j}N^{c,k}\right]\,, \label{RPVmodelI}
\end{eqnarray}
where the explicit indexes are flavor indexes of the relevant flavor gauge group. The RPV couplings for the light states originate from the mixing of the interaction eigenstates.
The three RPV couplings involving the vector-like quarks are $U^c_1D^c_2D^c_3$, $U^c_2D^c_1D^c_3$, $U^c_3D^c_1D^c_2$ and taking into account  the mixings, we get the effective $\Delta B=1$ R-parity violating couplings given in  Table~\ref{tab:couplingsmodel1}. 
In the Table
 we have omitted a factor $\lambda_q/(\lambda_d^2\lambda_{u})$ that is common to all the coupling and that we can take to be of order one.  The reason for the appearance of only one CKM factor is that we have chosen a basis where the down sector is flavor diagonal to start with and all CKM factors come from the up-sector. In the up sector, we get $U^{c,\prime}=V_{CKM} U^c$ and the mixing between heavy and light quarks is given by: $V^{\dagger}_{CKM}\frac{\lambda_u}{\hat{Y}_u}$.

\begin{table}
\begin{center} \begin{tabular}{||c|c|| }\hline
$\Delta B=1$ operator & strength \\[3pt]\hline
$u^cs^cb^c$ &$ {V_{ud} \, m_u\,m_s\,m_b}/{m^3_t}$\\[3pt]
$c^cs^cb^c$ &$ {V_{us}\, m_c\,m_s\,m_b}/{m^3_t}$\\[3pt]
$t^cs^cb^c$ &$ {V_{ub}\, m_t\,m_s\,m_b}/{m^3_t}$\\[3pt]
$u^cd^cb^c$ &$ {V_{cd}\, m_u\,m_d\,m_b}/{m^3_t}$\\[3pt]
$c^cd^cb^c$ &$ {V_{cs}\, m_c\,m_d\,m_b}/{m^3_t}$\\[3pt]
$t^cd^cb^c$ &$ {V_{cb}\, m_t\,m_d\,m_b}/{m^3_t}$\\[3pt]
$ u^cd^cs^c$ &$ {V_{td}\, m_u\,m_d\,m_s}/{m^3_t}$\\[3pt]
$ c^cd^cs^c$ &$ {V_{ts} \,m_c\,m_d\,m_s}/{m^3_t}$\\[3pt]
$ t^cd^cs^c$ &$ {V_{tb}\, m_t\,m_d\,m_s}/{m^3_t}$\\[3pt]
\hline
\end{tabular} \end{center}
\caption{Predictions for the $\Delta B=1$ RPV couplings in the gauged flavor model I. The strength of the coupling is given up to a factor $\lambda_q/(\lambda_d^2\lambda_{u})$ that is a parameter of the model and can be order one.}
\label{tab:couplingsmodel1}
\end{table}

\subsection{Model II}
The Model II is also based on the same electroweak gauge group and same matter content as Model I. However the flavor group is chosen as the $$SU(3)_{V,q}\times SU(3)_{V,\ell}$$ that are the diagonal combination of the $SU(3)_{Q_{L}}$ and $SU(3)_{Q_{R}}$ and the $SU(3)_{\ell_{L}}$ and $SU(3)_{\ell_{R}}$, respectively. 
This requires obvious changes in the assignment of flavor multiplet structure for all the fermions.
\begin{table}
\small
\begin{center} \begin{tabular}{|| l | c  c c c | c c  || }\hline\hline
& $SU(3)_c$  & $SU(2)_L$ & $SU(2)_R$ & $U(1)_{B-L}$ &  $SU(3)_{V,q}$ 
  & $SU(3)_{V,\ell}$  \\ [3pt]
\hline \hline
 \rowcolor{verdino} $Q$ & 3 & 2 & & $\frac{1}{3}$ & 3 &  \\ [3pt]
\rowcolor{verdino}  $Q^c$ & $3^{*}$& & 2 & $-\frac{1}{3}$ &  $3^*$ &    \\ [3pt]
  \rowcolor{giallino}   $U$ & $3$ & & & $\frac{4}{3}$    & $3^*$ &   \\ [3pt]
 \rowcolor{giallino}    $U^c$ & $3^{*}$ & & & $-\frac{4}{3}$   & 3 &   \\ [3pt]
 \rowcolor{giallino}    $D$ & 3 & & & $-\frac{2}{3}$  &  $ 3^*$ &   \\ [3pt]
 \rowcolor{giallino}    $D^c$ & $3^{*}$ & & & $\frac{2}{3}$  & $3$ &    \\ [3pt]
\hline
\rowcolor{verdino}  $L$  & & 2 & & $-1$ &   & 3    \\ [3pt]
\rowcolor{verdino}  $L^c$ & & & 2 & $1$ &   & $3^*$ \\ [3pt]
 \rowcolor{giallino}    $E$ & & &   & $-2$ & &   $3^*$ \\ [3pt]
 \rowcolor{giallino}    $E^c$ & & &   & $2$   &  &$3$   \\ [3pt]
 \rowcolor{giallino}    $N$ & & &   & 0  &    & $3^*$ \\ [3pt]
 \rowcolor{giallino}    $N^c$ & & &   & 0   & & $3$    \\ [3pt]
\hline
  $\chi,\bar{\chi}$ && 2 & & $\pm 1$ &   &  \\ [3pt]
  $\chi^c,\bar{\chi}^c$& & & 2 & $\mp 1$   &  & \\ [3pt]
  $\Delta_u$ &  &  & &  & $ 6$ &  \\ [3pt]
  $\bar{\Delta}_u$ & & &  &  & $ {6^*}$  & \\ [3pt]
  $\Delta_d$ &  &  & &  & ${6}$ &     \\ [3pt]
   $\bar{\Delta}_d$ & & &  &  & ${6^*}$    & \\ [3pt]
  $\Delta_\ell$ & &  & &  & &  $6$\\[3pt]
  $\bar{\Delta}_\ell$  & & & &  & & ${6^*}$\\[3pt]
$\Delta_\nu$ &  & & & & &  6\\[3pt]
$\bar{\Delta}_\nu$ & & & & &  & ${ 6^*}$\\[3pt]
\hline
\end{tabular} \end{center}
\caption{Model II matter content and transformation properties of the fields. The lines in green contain the MSSM quarks and leptons superfields, the line in yellow are the partner states. The electroweak and flavor symmetry breaking sector is given in the lines with white background.}
\label{tab:modelII}
\end{table}

In the Higgs sector, the flavons $Y_{u,d}$ are replaced by flavor group sextets $\Delta_q (1,1,0, 1, 6,1)$ and $\Delta_\ell (1,1,0,1,1,6)$ {supplemented by} their conjugate fields $\Delta^c_q$ and $\Delta^c_\ell$. {The representations under the gauge symmetry of the model
$$SU(2)_L\times SU(2)_R\times U(1)_{B-L}\times SU(3)_c\times SU(3)_{V,q}\times SU(3)_{V,\ell}\,$$
are given by the numbers in the parenthesis next to each field.}
 {To the best of our knowledge} such a gauged flavor model has not been discussed in the literature. Using the particle content in Table~\ref{tab:modelII}, we write down the RP conserving part of the superpotential of the model that gives mass to the fermions. In the quark sector, we have
\begin{eqnarray}
W_{II}&=&\frac{\lambda_u}{M}\left[[Q_i\chi U^c_j \bar{\Delta}^{ij}_u+Q^c_{i}\chi^c U_{j}\Delta_{u}^{ij}\right]+ M_1UU^c+ \\ \nonumber
&+&\frac{\lambda_d}{M}\left[Q_i\bar{\chi} D^c_j \bar{\Delta}^{ij}_d+Q^c_{i}\bar{\chi}^c D_{j}\Delta_{d}^{ij}\right]+ M_2DD^c
\end{eqnarray} 
and a similar part for leptons, which we do not display. {As in model I, for sake of simplicity we assumed L-R parity to be unbroken. The breaking of L-R parity does not change  our results qualitatively.} We remark that in this case the flavor gauge group is such that a mass term $UU^{c}$ and a similar term $DD^{c}$ can be written without breaking the the flavor gauge symmetry.} We also point out that since $\Delta_{u,d}$ are singlets of the electroweak gauge group, in principle there could also be couplings of type $Q_i\chi U^c_j \bar{\Delta}^{ij}_d$ and $Q_i\bar{\chi} D^c_j \bar{\Delta}^{ij}_u$; however, we can take appropriate linear combination of $\Delta_{u,d}$ and recast the couplings in the form as above by renaming the fields.
We also assume that minimization of the potential for the flavons $\Delta_{u,d}$ leads to a VEV pattern as follows:
\begin{eqnarray}
\langle \Delta_u\rangle = \langle \bar\Delta_u \rangle = V_{CKM}\cdot{\rm diag}(a_u, a_c, a_t)\cdot V^{\dagger}_{CKM},\quad \langle \Delta_d \rangle =\langle \bar\Delta_d \rangle =  {\rm diag}(a_d, a_s, a_b)\,.
\end{eqnarray}
Diagonalizing the resulting heavy-light quark mass matrices, we find that
\begin{eqnarray}
m_{u,c,t}\simeq \frac{\lambda^2_u v_Lv_R(a^2_{u,c,t})}{M^2M_1}, \quad m_{d,s,b}\simeq \frac{\lambda^2_d v_Lv_R(a^2_{d,s,b})}{M^2M_2} \label{quarkmassesMtwo} \,,
\end{eqnarray}
where, at variance with model I, the masses $M_{1,2}$ are now in principle unrelated to the flavon VEV.
The mixing between the heavy and the light quarks is given by two angles per generation: $\theta_{i L,R}$.  They are given by:
\begin{eqnarray}
\theta^{u,d}_{i, L, R} \simeq \frac{\lambda_{u,d}v_{L,R}(a_{u_i,d_i})}{MM_{1,2}}\,
\end{eqnarray}
and, since $v_R\gg v_L$, $\theta_R \gg \theta_L$, we get $$\theta^{u,d}_{i,R}\simeq \sqrt{\frac{m_{u_i,d_i}v_R}{m_t M_{1,2}}}\,.$$

In order to discuss the RPV couplings in the model, we write down the basic couplings prior to symmetry breaking:
\begin{eqnarray}
W_{RPV}=\lambda^{\prime\prime}[U^cD^cD^c+UDD]+\lambda (LLE^c+L^cL^cE)+\tilde\lambda (NNN+N^cN^cN^c)\,.\label{RPVmodeltwo}
\end{eqnarray}
Taking into account the mixings between the heavy and light quarks and using the formula for the quark masses eq.~(\ref{quarkmassesMtwo}), we find the generic form for the RPV in the light quark sector to be:
\begin{eqnarray}
W^{eff}_{RPV}~=~\lambda^{\prime\prime} y \epsilon_{jlm}\sqrt{\frac{m_{u_k}m_{d_l} m_{d_m}}{m^3_t} }u^c_jd^c_2 d^c_3 V^{CKM}_{kj}\,,
\end{eqnarray}
where $y=\sqrt{ \frac{v^3_R}{M^2_2M_1}}$  is a common overall factor for all the couplings that is expected to be smaller than one but presumably not much smaller. We will assume it to be $\sim 0.1$ in what follows.
Note the weaker (square root) dependence on the quark masses. The detailed RPV couplings involving different generations in this case is as in Table~\ref{tab:couplingsmodel1} with the proviso that the quark mass dependence is the square root of those in the Table~\ref{tab:couplingsmodel1} .

We reemphasize the point that the main reason for ``safe'' R-parity violation without proton decay problem in these models is the presence of separate horizontal symmetries for the quarks and leptons.
When this model is grandunified to an $SU(5)\times SU(5)$ group as in \cite{GUT55}  the quarks and leptons get unified and the two separate horizontal groups $SU(3)_{q,\ell}$ 
merge to one $SU(3)_H$ group. This results in an extra RPV term in the superpotential of the form $QLD^c$ and the proton decay problem resurfaces.

\section{Phenomenology of RPV couplings in gauged flavor \label{pheno}}
Our results have interesting phenomenological implications for collider searches for supersymmetry as well as for low energy baryon number violation where baryon number changes by two units.

\subsection{$\Delta B=2$ and gauged flavor RPV}
\begin{figure}
\begin{center}  \includegraphics[width=0.86\linewidth]{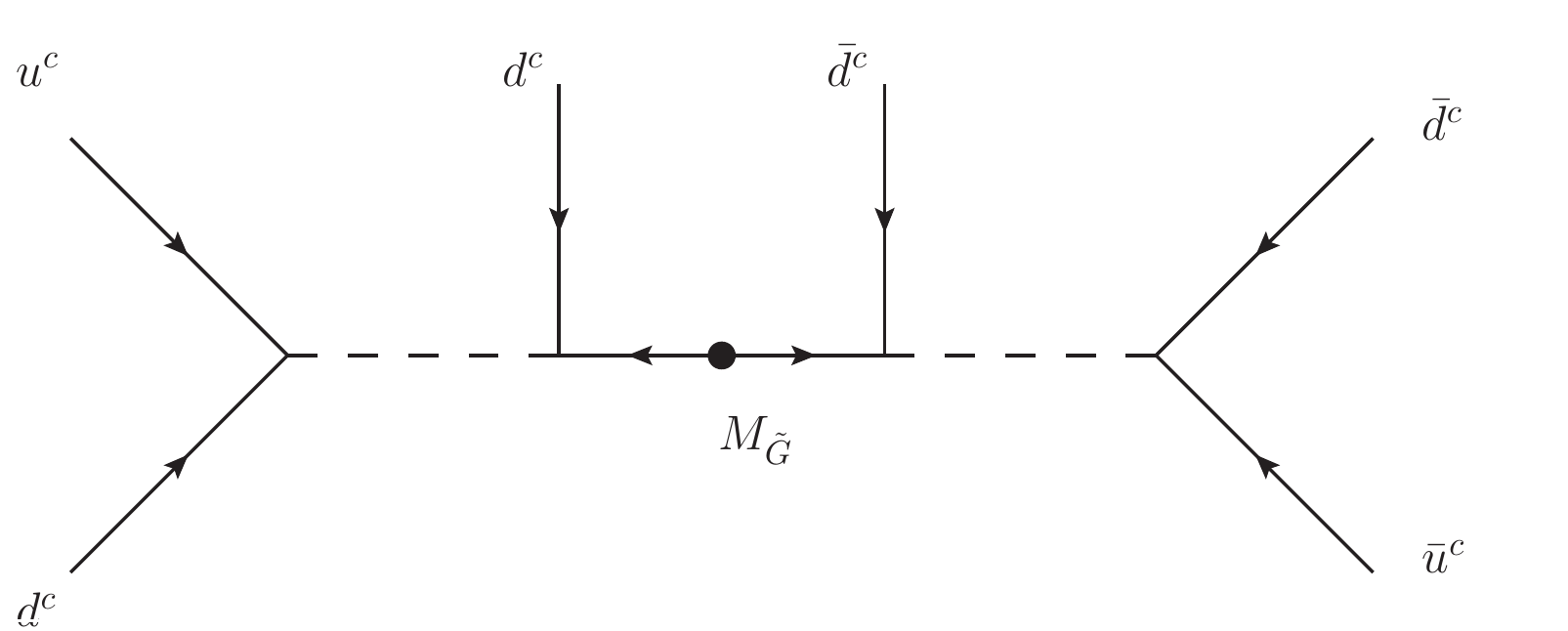}\end{center}
  \caption{The tree level diagram for$\Delta B=2$ process such as neutron-anti-neutron oscillation due to R-parity violating interaction. In this diagram when two external $d^c$'s are replaced by two $s^c$'s it gives rise to the double proton decay $pp\to K^+K^+$.\label{fig:nnbardiag}}
\end{figure}

The $\Delta B=1$ RPV couplings of model I and II after supersymmetry breaking lead to $\Delta B=2$ processes via the Feynman graph in Fig.~\ref{fig:nnbardiag}, where two outgoing fermion fields can either be $d^c$'s or  $s^c$'s. The $s^c$-term arises from the RPV interaction the $u^cd^cs^c$ term in the Table~\ref{tab:couplingsmodel1}, and it leads to the double nucleon decay process $pp\to K^+K^+$\cite{glashow} on whose lifetime there is a lower limit of $1.7\times10^{32}$~yrs~\cite{miura}. In our model I, we find the amplitude for this process to be:
\begin{eqnarray}
A_{pp\to K^+K^+}\sim \left(  \frac{V_{td} m_um_dm_s}{m^3_t}\right)^2 \frac{1}{M^4_{\tilde{q}} M_{\tilde{G}}}\,,
\end{eqnarray}
where $M_{\tilde{G}}$ is the gluino mass and $M_{\tilde{q}}$ is the relevant squark mass. This amplitude for model I is $\sim 10^{-45}$ GeV$^{-5}$ which is unobservable. For model II, due to the square root dependence, the life time for this process can be as large as $10^{34}$ years for a reasonable choice of the parameters ($v_R, M$, ...) and thus observable in the next round of proton decay search experiments.

The dominant contribution to neutron anti-neutron oscillation arises from the same Fig.~\ref{fig:nnbardiag}, where the RPV vertex on each side involves $u^cd^cb^c$ and in the dashed line, the $\tilde{b}^c$ has been converted to a $\tilde{d^c}$~\cite{babu}.  { In specific scenarios of SUSY breaking the $\tilde{b^c}-\tilde{d^c}$ mixing factor is given by~\cite{babu}
$\delta_{13,RR}\sim\frac{ \lambda^2_t}{8\pi^2}\frac{(3m^2_0+A^2_0)}{ m^2_{0}+8M^2_{1/2}}(V^*_{tb}V_{td})\ln(M/v)\simeq 2\times 10^{-4}$}. {Combining this with the RPV couplings in the models,  either $\frac{V_{cd}m_um_dm_b}{m^3_t}$ in model I or $\sqrt{\frac{V_{cd}m_um_dm_b}{m^3_t}}$ in model II,} leads to unobservable oscillation times ($10^{25}$ sec. in model I and $10^{15}$ sec. in model II) for neutron-anti-neutron oscillation. {The quantity $\delta_{13,RR}$ is allowed to be significantly larger by model-independent bounds~\cite{deltas}, however the oscillations are not observable even with this more generous source of flavor violation.}

\begin{figure}
\begin{center}  
\includegraphics[scale=0.75]{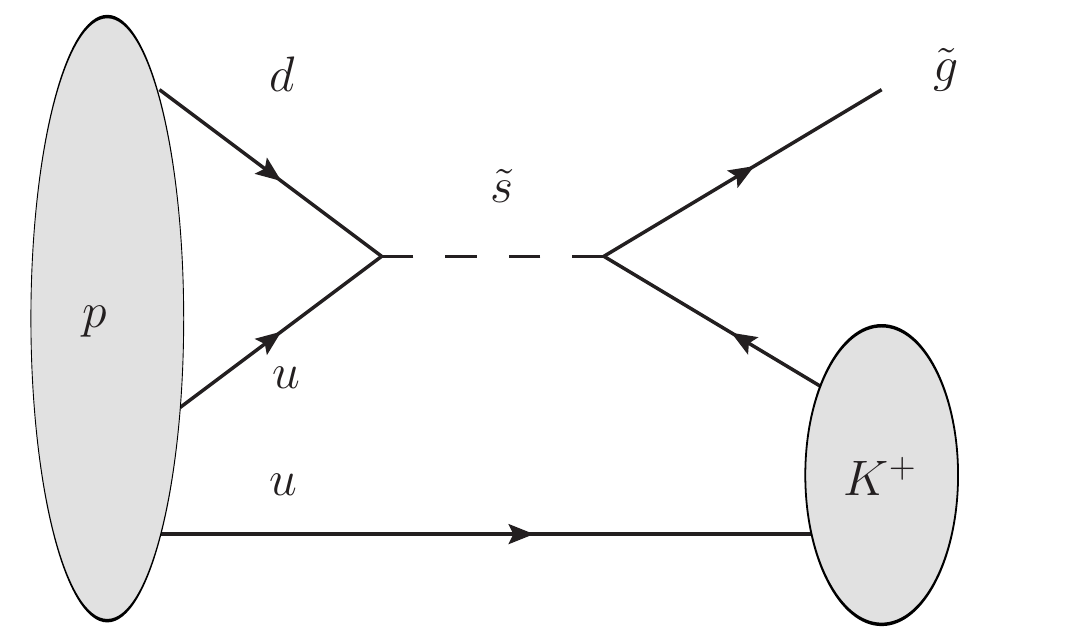}\includegraphics[scale=0.75]{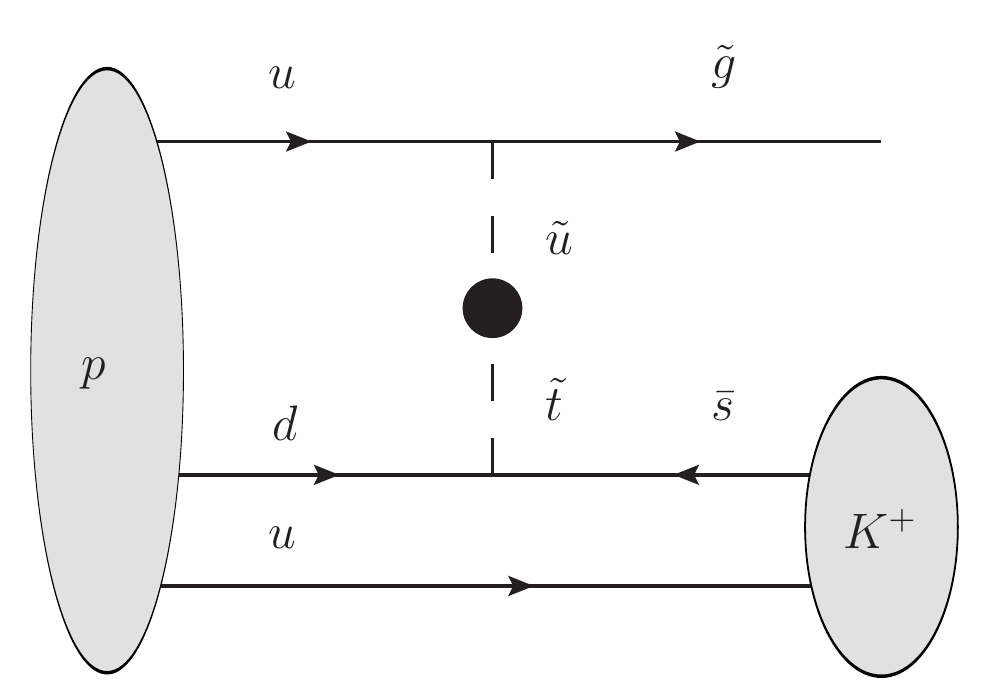}
\end{center}
  \caption{The tree level diagrams for $p\to K^+\,\tilde{g}$ decay due to R-parity violating interactions.\label{fig:pdecay}}
\end{figure}

An interesting property of this class of models  is that they  lead to an exotic decay mode for the proton if the gravitino is lighter than the proton, which can be true in gauge mediated susy breaking (GMSB) type models~\cite{Choi:1997ye}. The Feynman diagrams for this process is shown in Fig.~\ref{fig:pdecay}. The amplitude for this can be estimated roughly to be:
\begin{eqnarray}
A_{p\to K^++\tilde{g}}~\sim \frac{\lambda^{\prime\prime}}{M^2_{\tilde{q}}m_{3/2}M_P}
\end{eqnarray}
{For the model I we get that the decay is dominated by third generation squarks, putting in the values of the $t^c d^c s^c$ type RPV coupling i.e. $\lambda^{\prime\prime}\simeq V_{tb}\frac{m_d m_s}{m^2_t}$ (setting $\lambda_{u,d}\sim 1$ ), proton life time for a keV gravitino is $\tau_{p\to K^++\tilde{g}}\sim 10^{33}\left(\frac{m_{3/2}}{1{\rm KeV}}\right)^2 \left( \frac{0.01}{\delta_{13}^{u}} \right)^{2}$ yrs (for TeV squark masses). The current limits  on $p\to K^+\nu$ should apply. For $\tau_{p\to K^{+}\nu}>3.3\times10^{33}$~yrs~\cite{Kobayashi:2005pe} one should have $m_{3/2} \gtrsim $~KeV assuming $\delta_{13}^{u}\sim0.01$ as in gravity mediation schemes for the mediation of SUSY breaking. Turning to the model II where the RPV coupling goes like square root of quark mass products, under the same assumptions as above the current proton life time limits will limit $m_{3/2} \gtrsim $ MeV.}

\subsection{Leptonic RPV}
As we saw in eq.~(\ref{RPVmodeltwo}), there are pure leptonic RPV couplings allowed by the gauge symmetry.   They are of the form $LLE^c+L^cL^cE$. These couplings at the one loop level lead to equal Majorana mass for all neutrinos. Their magnitude is given by:
\begin{eqnarray}
m_{\nu_i}\simeq \frac{A m^2_{\tau}}{16\pi^2 v^2_R \lambda^2}\,.
\end{eqnarray}
For $\lambda\sim 1-3$, this can be in the eV range for $A\sim $ GeV or so. This couplings also leads to decay of sneutrinos to $\ell^+_i\ell^-_j$. We defer the discussion of neutrino masses in this model to another paper.

\subsection{LHC and RPV}
We have outlined the possible patterns of the RPV couplings that can arise depending on the gauged flavor symmetry that one considers. In particular it emerged that the overall scale of the couplings depends on free parameters, as for instance remarked commenting about Table~\ref{tab:couplingsmodel1}. Despite this parametric uncertainty on the overall scale of the RPV couplings, the models summarized in Tables~\ref{tab:charges}, \ref{tab:modelone}, and~\ref{tab:modelII} have characteristic couplings ratios. For instance the RPV couplings of the $bs$ and $bd$ quark bilinear with the stop are much less hierarchical in model II than in the other models. This has potentially observable consequences. For instance one can estimate  the branching ratio of a stop LSP into a $b$-jet and a light jet
\begin{eqnarray}
{\rm Br}\left(\tilde{t} \to b q\right) &\simeq& \frac{(m_b m_s V^2_{ub} )+(m_b m_d V^2_{cb} )}{  (m_d m_s V^2_{tb}) + (m_b m_s V^2_{ub} )+(m_b m_d V^2_{cb} )}  \simeq 0.14 \quad  \textrm{in model II,}\\
{\rm Br}\left(\tilde{t} \to b q\right) &\simeq& \frac{(m_b m_s V_{ub} )^{2}+(m_b m_d V_{cb} )^{2}}{  (m_d m_s V_{tb})^{2} + (m_b m_s V_{ub} )^{2}+(m_b m_d V_{cb} )^{2}  } \simeq 0.99 \quad \textrm{in model I,}
\end{eqnarray}
for $m_{b}=4\gev$, $m_{s}=0.1\gev, m_{d}=6\mev$ and $V_{ub}=\lambda^{3}$, $V_{cb}=\lambda^{2}$, $V_{tb}=1$, $\lambda=0.22$.

The effect is quite sizable and  measurable. In fact the difference between the prediction of the two models is much larger than the uncertainties  due to the uncertainty on the theory parameters.  In case the stop manifests itself in signatures with associated charged leptons~\cite{Brust:2012yq}, the distinction of the decay modes with heavy flavors from those in light quarks seems attainable. In this, and in all the cases where both the decays $\tilde{t}\to bq$ and $\tilde{t}\to qq$ give rise to observable signals, the uncertainty of the $b$-tagging and signal selection efficiencies needs to be well under control to measure this branching ratio with sufficient accuracy.
Depending on the spectrum of the few lightest SUSY particles, the measurement could be more challenging because in general one is not guaranteed to have  extra leptons associated to the production of the stops. This is the case for instance when the stop is direclty produced from QCD interactions $pp\to \tilde{t}\tilde{\bar{t}}$. In absence of an associated lepton the $bq$ decay mode of the stop can still be observed \cite{Franceschini:2012fu}, but the decay into light jets may not be observable due to a large background from QCD multi-jet production \cite{Evans:2012oq}. In this case the ${\rm Br}(\tilde{t}\to bq)$ may be inferred from the absolute rate, but given the uncertainty inherent in the QCD production mechanism of the stops we expect a not very significant discrimination between the two models.

Analogously to the decay of a stop LSP, if the sbottom is the LSP, the different hierarchy of the RPV couplings of model II will result in a reduced ${\rm Br}(\tilde{b}\to tq)$ compared to the prediction of the other models. The expected reduction of the branching fraction is similar to that for the stop LSP case.

The fact that the strongest RPV couplings are the $t^cs^cb^c$, $t^cd^cb^c$ and $t^cd^cs^c$ couplings also have implications for the possible phenomenology of a light gluino. In fact if the  gluino is the LSP, or it is light, its decay will result in the overall production of three quarks $$\tilde{g}\to q q q $$ whose flavor depends on the strength of the RPV couplings and the actual squark spectrum. Unless the sbottom and the stop are much heavier than the other squarks, the final state of the gluino decay is expected to have some heavy flavors.

A resonance that closely resembles the RPV gluino, produced in pair and decaying into three light flavored jets, has been searched  by the ATLAS collaboration~\cite{ATLAS-Collaboration:2012fl}.  In Ref.~\cite{Allanach:2012ai} other searches from the ATLAS and CMS collaboration have been reinterpreted in terms of RPV SUSY gluino with light sbottom or stop squarks. The most important reinterpreted searches are those that look at a final state with two leptons of same charge, jets and missing transverse momentum~\footnote{The presence of same-sign leptons in the RPV SUSY signal is largely due to the Majorana nature of the gluino. For a Dirac gluino same-sign leptons would be much less abundant.}. Although not used yet in the current searches, flavor information about the jets could be used to probe the structure of the RPV couplings. However we remark that  the rates of each flavor combination of the gluino decay depends both on the squark spectrum and on the RPV couplings. Therefore to pinpoint a structure in the RPV couplings some other data must provide information on the squark masses.

As said above, the overall scale of the RPV couplings depends on free parameters of the model. Therefore it is hard to make a definitive statement on the life-time of the LSP that decays through RPV couplings. Despite this arbitrariness, it seems natural to imagine that, for typical choices of these parameters,  model II has a larger overall scale of the RPV couplings. This is due to the different power of the ratio of masses that are relevant for the couplings of model II.
For this reason we expect that in Model II it is not typical to get displaced decays of the LSP, which otherwise would be a potentially very effective way to reject many SM backgrounds. For the same reason, the interesting phenomenology of mesino oscillations and the bounds discussed in 
Refs.~\cite{Berger:2012yq,Evans:2012oq} should not apply for model II.

\section{Conclusion}
In summary, we have pointed out that supersymmetric versions of a class of gauge flavor models provides a realization of ``natural'' R-parity violation which unlike MSSM are not threatened by catastrophic proton decay problem. The strengths of RPV interactions are found to have interesting dependence on quark masses, which are different from those in minimal flavor violating models. {In one class of models we find linear products of quark masses characterizing the strengths of RPV interactions whereas in the another class the RPV couplings have an extra square root, making them less hierarchical.} We discuss several phenomenological implications of this model e.g. double proton decay to two strange mesons, proton decay to superlight gravitinos as well as some LHC implications of the hierarchical nature of the RPV interactions.

\section*{Acknowledgments} This  work 
is supported by the National Science Foundation Grant No.~PHY-0968854. The work of R.~F. is also supported by the National Science Foundation Grant No.~PHY-0910467, and by the Maryland Center for Fundamental Physics.
R.~F. thanks Daniel Stolarski for discussions.


\end{document}